\documentstyle[aps,multicol,epsfig]{revtex}
\input epsf
\begin{document}
\title {In-plane gate single-electron transistor in 
Ga[Al]As fabricated by scanning probe lithography}
\author{ S. L\"uscher, A. Fuhrer, R. Held, T. Heinzel, and K. Ensslin}
\address{Solid State Physics Laboratory, ETH Z\"urich, 
CH-8093 Z\"urich, Switzerland}
\author{W. Wegscheider $^{*}$$^{+}$}
\address{$^{*}$Walter Schottky Institut, TU M\"unchen, 85748 Garching, 
Germany}
\address{$^{+}$ Institut f\"ur Angewandte und Experimentelle Physik, 
Universi\"at Regensburg, 93040 Regensburg, 
Germany}
\maketitle
\begin{abstract}
A single-electron transistor has been realized in a Ga[Al]As 
heterostructure by oxidizing lines in the GaAs cap layer with an atomic 
force microscope. The oxide lines define the boundaries of the 
quantum dot, the in-plane gate electrodes, and the contacts of the 
dot to source and drain. Both the number of electrons in the dot 
as well as its coupling to the leads can be tuned with 
an additional, homogeneous top gate electrode. Pronounced Coulomb 
blockade oscillations are observed as a function of voltages applied 
to different gates. We find that, for positive top-gate voltages, the 
lithographic pattern is transferred with high 
accuracy to the electron gas. Furthermore, the dot shape does not change significantly 
when in-plane voltages are tuned.
\end{abstract}
\begin{multicols} {2}
\narrowtext
Single electron transistors (SETs) \cite{Grabert92} are currently the 
subject of intense 
research activities.  They can be used as devices 
which rely on the discreteness of the electron charge, for example 
highly sensitive electrometers \cite{Lafarge91,Yoo97,Wei97,Koltonyuk99}, 
or single electron pumps as a possible current standard  
\cite{Martinis94}. 
Furthermore, SETs defined in a two-dimensional electron gas 
(2DEG) of a semiconductor heterostructure are also known as "quantum dots", 
and can be seen as tunable artificial atoms 
with interacting electrons. The interplay between Coulomb interactions 
and quantum 
size effects has been widely studied in such structures.\cite{Kouwenhoven97}

Generally 
speaking, an SET consists of a conducting island, which is weakly 
coupled to two reservoirs via tunnel barriers. In order to be able to tune 
the electrochemical potential of this island, a nearby gate 
electrode is coupled capacitively to it. In recent years, a 
variety of fabrication methods for tunable semiconductor quantum dots has 
been reported, each having its particular strengths and weaknesses. 
The most common scheme consists of patterning metallic top gate 
electrodes by electron beam lithography.\onlinecite{Kouwenhoven97} 
By applying negative voltages 
to these gates, the quantum dot is induced in the electron gas. 
Its shape and electron density as well as its coupling to the 
leads can be tuned over wide ranges. The disadvantage of this method is that due 
to the large lateral depletion lengths of the order of 100nm,  the 
dot shape deviates significantly from the top gate pattern. 
Furthermore, changing the number of electrons inside the dot changes 
its shape as well, which is reflected experimentally in gate-voltage 
dependent capacitances. Patterning the dot and in-plane gate electrodes by 
wet chemical etching results in quite similar advantages and disadvantages. 
\cite{Pothier93,Wang96} Other 
schemes of fabricating  quantum dots are focused laser beam-induced 
doping,\cite{Baumgartner97} and focused Ga ion beam implantation in combination with 
top gates\cite{Fujisawa96}. Here, the lateral depletion is smaller. However, it is known 
that scattering at edges defined by such implantation techniques is highly 
diffusive,\cite{Thornton89} 
thus reducing, or even destroying, ballistic transport through the 
dot.\\
\begin{figure} 
\centerline{\epsfxsize=8.0cm \epsfbox{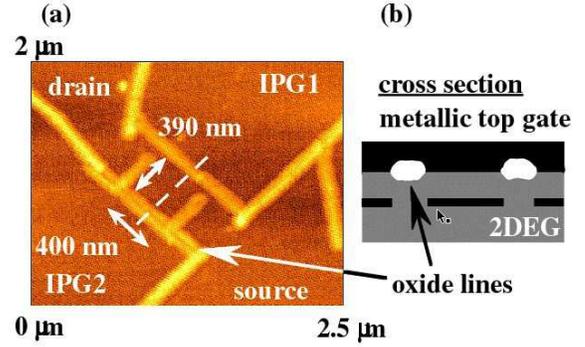}}	
\caption{ Fig. 1: (a): Scanning force micrograph of the oxide pattern that 
defines the quantum dot, which is coupled to source and drain 
via small gaps in the oxide lines. The dot
potential can be tuned using the electrodes IPG1 or IPG2. The 
picture shows the sample before the top gate metallization. The 
dashed line indicates the line of the  cross section  scheme in (b). 
The oxide lines deplete the electron gas underneath and separate the 
dot from IPG1 and IPG2. The sample is homogeneously covered by a Ti/Au 
top gate.}
\end{figure}  
Here, we report the realization of quantum dots in a 
$GaAs-Al_{x}Ga_{1-x}As$ (x=0.3) heterostructure by local oxidation 
 of the semiconductor surface
with an atomic force microscope (AFM). \cite{Held98} Clear Coulomb blockade of the electron 
transport is observed. It is demonstrated that by combining top gates 
with in-plane gates, the lithographic shape of the dot can be transferred into 
the electron gas with high accuracy. 
Furthermore, by applying voltages to the in-plane gates, the occupation number 
of the dot can be tuned over 
wide ranges by leaving its shape basically unchanged.\\ 
The 2DEG in our samples is located $34 nm$ below 
the surface. It has a sheet density of $5.0\cdot10^{15}m^{-2}$ and a mobility 
of $92 m{^2}/Vs$ at a temperature of T=0.1 K. 
The coarse lateral structure of the sample is pre-patterned by optical 
lithography. A Hall bar is defined by wet chemical etching, and  Ni-AuGe Ohmic contacts are
alloyed into the heterostructure. Then, the dot is defined by local 
oxidation (LO) of the GaAs cap layer 
(Fig. 1a). Oxide lines (typical width 100nm and height 8nm, 
respectively) deplete the 
electron gas underneath \onlinecite{Held98}, 
separate the quantum dot from the in-plane gates (IPG1 and IPG2) and define the 
connections to source and drain. In a final fabrication step, the 
sample is covered with a homogeneous layer of 100nm Au on top of 5nm 
Ti, acting as a top gate (Fig. 1b). Recently, Sasa et al. 
\cite{Sasa99} reported the observation of Coulomb blockade in a quantum 
dot defined by combining LO on InAs/AlGaSb heterostructures with a 
selective wet chemical etching technique. Our technique, which works in 
Ga[Al]As, does not need an etch step and can be combined with additional 
top gates, which may be laterally patterned. \\
DC conductance measurements have been carried out in a $^{3}He/^{4}He$ 
dilution refrigerator with a base temperature of 50mK. No leakage 
currents $(I\leq 1pA)$ between top gate and 2DEG could be detected for top gate 
voltages $-2V\leq V_{tg}\leq+1V$. The depleted regions under the oxide 
lines have typical breakdown voltages of $\pm 400mV$ which, however, 
depend on $V_{tg}$.\cite{Held99} Applying voltages to the top gate changes not 
only the occupation number of the dot, but also its coupling to the leads. 
We have studied two different realizations of such quantum dots.
In type 1, the dot is 
separated from source and drain via two highly 
insulating tunnel barriers when the top gate is grounded. The tunnel barriers are defined 
by a gap of 
about $20nm$ in the two 
oxide lines that separate the dot from source and drain. By applying positive voltages to the top gate, 
the  tunnel barriers can be opened. 
In type 2,  the dot is coupled to source and drain via open quantum point 
contacts  when the top gate is grounded. This is achieved by keeping 
the gaps in the oxide lines at $\approx 50nm$ (Fig. 1a). Here, negative top gate 
voltages close the quantum point contacts.\\
In Fig. 2, conductance 
measurements on a type 1 dot are shown. Using a bias voltage of 
$20\mu V$, a current through the dot can be detected for 
$V_{tg}¥\geq 360mV$, and Coulomb blockade oscillations  
as a function of $V_{tg}$ can be observed for $380 mV\leq V_{tg}\leq 
400mV$ (Fig. 2a). An average  peak separation of $\Delta V_{tg}=240\mu V$ is 
found, corresponding to a capacitance between top gate and dot 
$C_{dt}$ of $670aF$. We modeled $C_{dt}$ by a parallel plate 
capacitor. Using the distance between top gate 
and the 2DEG as the separation between the plates, and  a 
dielectric constant of $\epsilon =13$, we 
find an electronic dot area of $ 2\cdot 10^5nm ^{2}$, which indicates 
that the 
lateral depletion length is clearly less than $20nm$ at the oxide 
lines, if about $400mV$ are applied to the top gate. This is in 
agreement with our 
earlier investigations on the depletion lengths of nanostructures 
defined by local oxidation. \cite{Held99} 
Hence, the size and shape of our 
quantum dot resemble much more closely the lithographic pattern than that defined 
solely by top gates (like in conventional split-gate structures) do. 
\begin{figure} 
\centerline{\epsfxsize=8.0cm \epsfbox{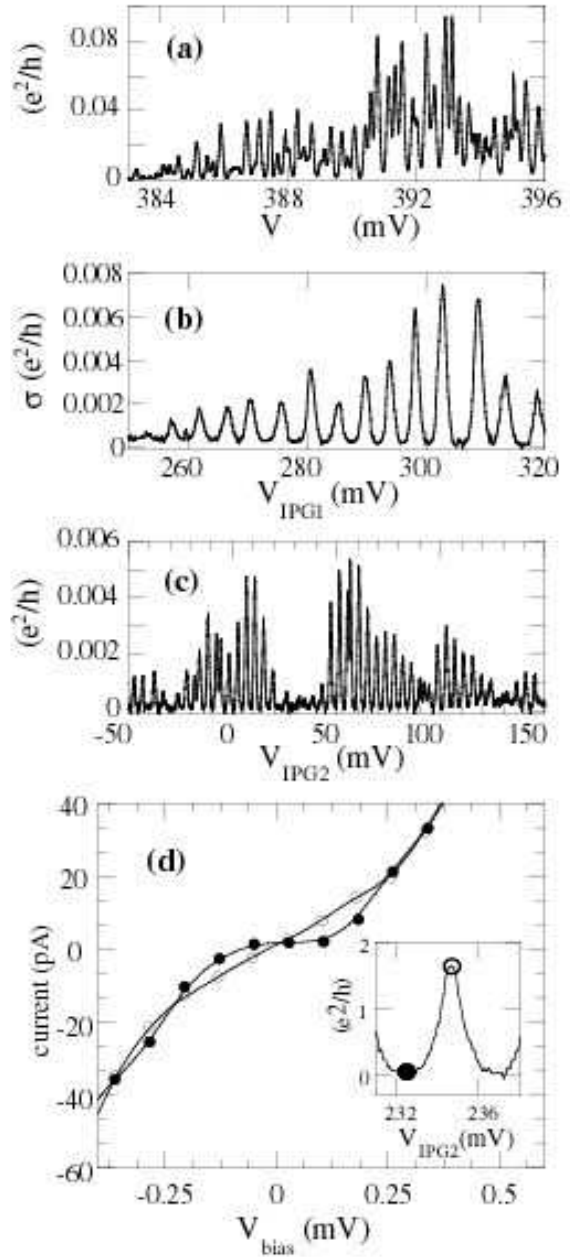}}	
\caption{ Fig. 2:  Conductance $\sigma$ through the type 1 dot as a function of the top 
gate voltage (a), the voltage applied to IPG1 (b) and IPG2 (c). 
The temperature was $T=120mK$
(d): I-V trace in the peak (open 
circles) and in the valley  (full circles) of the Coulomb resonance. 
The inset shows the Coulomb peak used for the I-Vs.}
\end{figure}
The tuning range for IPG2 corresponds to about 100 electrons (Fig. 2c), 
while the tuning range of IPG1 is only about 20 
electrons (Fig. 2b), which reflects the different influence of these gates on 
the tunnel barriers. From the average conductance peak separation as a function of 
the in-plane gate voltages, we determine the 
capacitances for IPG1 and IPG2 to be $33aF$ and $39aF$, 
respectively. 
 With a 3-dimensional numerical capacitance simulation we 
find excellent agreement between theoretical and experimental
values of the capacitances of the dot to all three 
gates \cite{capac}. \\
In Fig. 2d, I-V traces for two different IPG2 voltages are shown. The 
trace with the full circles,  taken in the valley of a Coulomb 
oscillation, shows a clear Coulomb gap. From a set of 
 I-V measurements over several Coulomb oscillation periods, we find a 
 Coulomb gap of  $410\mu eV$.\\
\begin{figure} 
\centerline{\epsfxsize=8.0cm \epsfbox{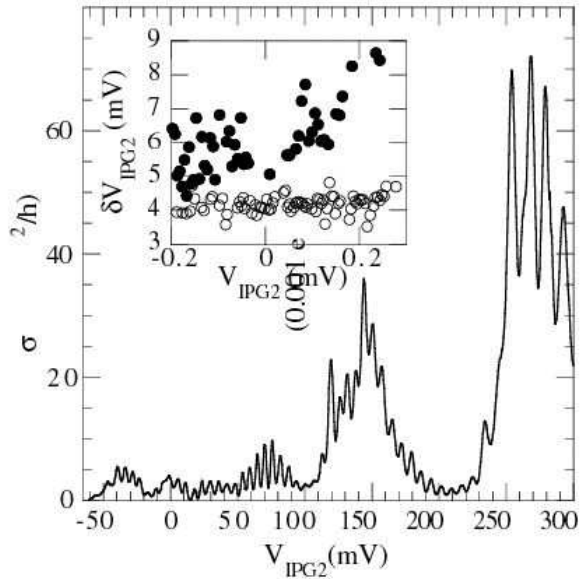}}	
\caption{Fig. 3: Conductance $\sigma$ through the type 2 dot. 
Superimposed to the Coulomb oscillations, transmission 
resonances are observed. In the inset we plot the periods of the conductance oscillations 
$\delta V_{g}$ as a function of  $V_{IPG2}$ 
obtained for the different dot types. The open circles 
represent the period of the peaks in the type 1 dot, which 
remains almost constant over the whole voltage range.  For comparison, the 
periods in $V_{IPG2}$, measured in a dot of type 2, i.e. under negative top gate 
voltages, are represented by the full circles. Here, the period increases significantly 
with increasing $V_{IPG2}$, since the capacitance between dot and gate 
changes.  }
\end{figure}  
In Fig. 3, we show the conductance through the type 2 dot. In addition to 
the Coulomb oscillations, one observes well-known superimposed transmission 
resonances. \cite{Heinzel94} In the case of this open dot, we have to 
apply $V_{tg}=-130 mV$ in order to close the 
quantum point contacts. It is obvious that, in this case, the Coulomb period 
increases with $V_{IPG2}$. In contrast, we observe in the type 1 dot, where we 
apply a positive top gate voltage,  an approximately 
constant period of about $4mV$ over wide ranges of $V_{IPG2}$.
 This different behavior is visualized in the inset of Fig. 3, where the 
 Coulomb peak separation as a function of $V_{IPG2}$ for both 
 dot types is shown.
 We interpret the constant peak separation in the type 1 dot (open 
 circles) as a consequence of the huge carrier 
 density due to the positive top gate voltage.  Therefore, the electric fields 
from the in-plane gates are screened very effectively 
and the dot keeps its shape over the whole  $V_{IPG2}$ range. This is 
not the case in the type 2 dot (full circles), where a negative top 
gate voltage is applied, which reduces the electron density in the 
system.\\
High quality SETs can be fabricated by a variety of methods. Our approach
leads to quantum dots that are highly tunable by in-plane as well as by top
gate electrodes. Furthermore the substrate as well as the electrodes can be
patterned by scanning probe lithography in two consecutive steps since
overlay accuracy is not a critical issue with an AFM. In particular we expect to be
able to realize dots containing few electrons where the charge and spin
degrees of freedom can be controlled with high accuracy.\\
In summary, we have shown that sub-micron size quantum dots can be 
patterned in Ga[Al]As heterostructures by combining local oxidation 
of the semiconductor surface with additional top gates, and Coulomb 
blockade can be observed. Operating the quantum dots with positive 
top gate voltages results in excellent transfer of the lithographic 
pattern into the electron gas. In addition, the dot keeps its shape 
while it is tuned by in-plane gates. Hence, our fabrication scheme is 
highly flexible and 
well suited for performing experiments in which the energy spectrum of 
quantum dots is studied.  \\
We wish to thank M. Bichler for growing the samples and G. Salis for his help with the sample preparation.
Financial support from the Schweizerischer Nationalfonds is 
gratefully acknowledged.\\

\end{multicols}

\begin{thebibliography}{99}
\bibitem{Grabert92} For a review, see Single Charge Tunneling, edited 
by H. Grabert and M. H. Devoret (Plenum, New York1992)
\bibitem{Lafarge91} P. Lafarge, H. Pothier, E.R. Williams, D. Esteve, 
C. Urbina, and M.H. Devoret, Z. Phys. B {\bf 85}, 327 (1991).
\bibitem{Yoo97} M.J. Yoo, T.A. Fulton, H.F. Hess, R.L. Willett, L.N. 
Dunkelberger, R.J. Chichester, L.N. Pfeiffer, and K.W. West, Science {\bf 
276}, 579 (1997). 
\bibitem{Wei97} Y.Y. Wei, J. Weis, K.v. Klitzing, and K. Eberl, Appl. 
Phys. Lett. {\bf 71}, 2514 (1997).
\bibitem{Koltonyuk99} M. Koltonyuk, D. Berman, N.B. Zhitenev, R.C. 
Ashoori, L.N. Pfeiffer, and K.W. West, Appl. 
Phys. Lett. {\bf 74}, 555 (1999).
\bibitem{Martinis94}J.M. Martinis, M. Nahum, and H.D. Jensen, Phys. 
Rev. Lett.  {\bf 72}, 904 (1994).
\bibitem{Kouwenhoven97} For a review, see L.P. Kouwenhoven, C.M. Marcus, P.L. McEuen, S. Tarucha, R.M. 
 Westervelt, and N.S. Wingreen, "Electron Transport in 
Quantum Dots",  in Mesoscopic Electron Transport,  Proceedings of a 
NATO Advanced Study Institute, 
 edited by L.P. Kouwenhoven, G. Sch\"on and L. L. Sohn, 
 (Kluwer,  Dodrecht, Netherlands, 1997), ser. E, vol. 345, pp. 105-214.
 \bibitem{Pothier93} H. Pothier, J. Weis, R.J. Haug, K. v. Klitzing, 
and K. Ploog, Appl. Phys. Lett. {\bf 62}, 3174 (1993).
\bibitem{Wang96} T. H. Wang and S. Tarucha, Appl. Phys. Lett. {\bf 69}, 
406 (1996).
\bibitem{Baumgartner97} P. Baumgartner, W. Wegscheider, M. Bichler, G. 
Schedelbeck, R. Neumann, and G. Abstreiter, Appl. Phys. Lett. {\bf 70}, 
2135 (1997).
\bibitem{Fujisawa96} T. Fujisawa and S. Tarucha, Appl. Phys. Lett. {\bf 68}, 
526 (1996).
\bibitem{Thornton89} T. J. Thornton, M. L. Roukes, A. Scherer, and B. P. 
 Van de Gaag, Phys. Rev. Lett. {\bf 63}, 2128 (1989).
 \bibitem{Held98} R. Held, T. Vancura, T. Heinzel, K. Ensslin, M. 
Holland and W. Wegscheider,
Appl. Phys. Lett. {\bf 73}, 262 (1998).
\bibitem{Sasa99} S. Sasa, T. Ikeda, K. Anjiki, and M. Inoue, Jpn. J. 
Appl. Phys. {\bf 38}, 480 (1999).
\bibitem{Held99} R. Held, S. L\"uscher, T. Heinzel, K. Ensslin and
 W. Wegscheider, Appl. Phys. Lett. {\bf 75}, 
1134 (1999).
\bibitem{capac} The simulation program can be downloaded 
from: ftp://rle-vlsi.mit.edu/pub/fastcap/
\bibitem{Heinzel94} T. Heinzel, S. Manus, D.A. Wharam, J.P. Kotthaus, G. 
B\"ohm, W. Klein, G. Tr\"ankle, and G. Weimann, Europhys. Lett.  {\bf 26}, 689  (1994).
\end{thebibliography}
\end{document}